# A SYSTEMATIC APPROACH TO CLEANING ROUTINE HEALTH SURVEILLANCE DATASETS: AN ILLUSTRATION USING NATIONAL VECTOR-BORNE DISEASE CONTROL PROGRAMME DATA OF PUNJAB, INDIA

Gurpreet Singh, Sree Chitra Tirunal Institute for Medical Sciences and Technology, India, drgurpreet.md.afmc@gmail.com

Biju Soman, Sree Chitra Tirunal Institute for Medical Sciences and Technology, India, bijusoman@sctimst.ac.in

Arun Mitra, Sree Chitra Tirunal Institute for Medical Sciences and Technology, India, arunmitra2003@gmail.com

**Abstract.** Advances in ICT4D and data science facilitate systematic, reproducible, and scalable data cleaning for strengthening routine health information systems. A logic model for data cleaning was used and it included an algorithm for screening, diagnosis, and editing datasets in a rule-based, interactive, and semi-automated manner. Apriori computational workflows and operational definitions were prepared. Model performance was illustrated using the dengue line-list of the National Vector Borne Disease Control Programme, Punjab, India from 01 January 2015 to 31 December 2019. Cleaning and imputation for an estimated date were successful for 96.1% and 98.9% records for the year 2015 and 2016 respectively, and for all cases in the year 2017, 2018, and 2019. Information for age and sex was cleaned and extracted for more than 98.4% and 99.4% records. The logic model application resulted in the development of an analysis-ready dataset that can be used to understand spatiotemporal epidemiology and facilitate data-based public health decision making.

## 1. INTRODUCTION

Routine Health Information Systems (RHIS) includes data that is collected at regular intervals from multiple health facilities including community-level public health centers, public and private hospitals, and other healthcare institutions (MEASURE Evaluation, 2021). These datasets provide information on health status, health services, and resources available for improving the health of populations. The strengthening of RHIS has emerged as a global as well as national agenda in numerous countries for data-driven decision-making. The processes involved in RHIS strengthening are thus looked at from a broader perspective beyond the data collection and entry processes. Harrison et al. suggest five pillars that form the basis of the simplified theory of change in strengthening routine health surveillance data for decision making: governance, people, tools, processes, and evidence (Harrison et al., 2020). The framework provided by World Health Organization to strengthen health systems includes health information as one among the identified attributes of a health system (World Health Organization, 2007). Also, the current existing initiatives such as the "Performance of Routine Information System Management" (PRISM) framework suggested by the Measure evaluation study group addresses many of the current challenges for improving data quality through the data life cycle. Some of the aspects involved in the data life cycle as suggested in the PRISM framework for evaluation of routine health information systems include behavioral challenges, environmental challenges, organizational challenges, and technological challenges (MEASURE Evaluation, 2021).

Good quality data is paramount to the success of health information systems. Generally, data is considered high-quality if it is "fit for [its] intended uses in operations, decision making and planning while representing the real-world constructs it" (Fadahunsi et al., 2019). Data quality of routine health information systems has been a subject of extensive research over the years. Availability of good quality data at timely intervals is critical to data-based public health decision-making (AbouZahr & Boerma, 2005). In addition to social, economic, political, and local contextual factors, multiple factors have been identified at each stage of the data life cycle which affect data quality. The quality of data collected is largely influenced by the level of work engagement, training, and perceived-self efficacy of the individual collecting data. Health system-related factors such as multiple communication channels, increasing variables for data entry, limited health infrastructure, and frequent changes in reporting formats are known challenges to good quality data (Aiga et al., 2008; Glèlè Ahanhanzo et al., 2014).

Advances in Information and Digital Technologies and data science approaches have potential in cleaning and extraction of information from routine large datasets. Routine health information datasets are prepared primarily for administrative and programmatic use. As a result, the data quality standards laid for monitoring data elements are thus bound to be defined differently when compared to those required for research-level datasets. This brings forward the need for data cleaning measures on raw routine datasets before use for research purposes. Inability to identify data anomalies efficiently leads to loss of information, high missing values, and inaccurate outcomes (Maïga et al., 2019; Van den Broeck et al., 2005). A systematic approach to data cleaning is recommended along with transparent documentation, however, there is a dearth of studies that explicitly disclose the steps followed and anomalies detected and corrected during data cleaning (Maina et al., 2017; Wilhelm et al., 2019). Further, it is essential to understand data cleaning as a systematic process rather than a one-time activity. The importance of data cleaning in the data lifecycle is crucial as the resultant data's quality would not only determine the robustness and generalizability but also allow for data linkage and sensible extrapolation of the study findings (Gesicho et al., 2020; Phan et al., 2020; Randall et al., 2013; Van den Broeck et al., 2005). Adopting a systematic approach to data cleaning would enable the researcher to find anomalies more efficiently and allow for reproducibility and transparency of the data lifecycle (Huebner et al., 2016).

The implications of open-source algorithms using technological advances on the future public health landscape are enormous. The volume of the data that flows through a health system is enormous and ever-increasing. Studies have documented that the data volume in the digital universe is doubling every two years (Oracle India, 2021). Further, data integrity and data consistency have been raised by many in context to the routine health information system (Smeets et al., 2011). Though a lot of light has been shed on data quality assurance and data quality control, these principles are yet to be translated into practice, especially in low-and-middle-income country settings including India. Evidence-informed data-based real-time decision-making by health program managers and data users require efficient data cleaning processes to extract information and knowledge from data. Manually, this process is not standardized, time and resource-intensive, and is often faced with manual omissions and commissions. The development of reproducible algorithms will enable efficient data cleaning on one hand and will provide solutions to numerous challenges which country is facing in terms of estimating the real-time burden of diseases, capacity building, rapid public health decision making, and thus enhanced prevention and control of diseases.

The use of open-source and reproducible algorithms will enable the generation of semi-automated mechanisms for data cleaning and provide transparency to the cleaning process followed. As it is important to study attributes related to the decision-makers such as data use culture, personal believes, and power relations in the organization to strengthen information systems, at the same time, it is prudent to look at the technological challenges when dealing with routine health information systems in the 21st century. The data science approach is useful in achieving this humongous task efficiently and scientifically. This is especially relevant as much of the data collected through the routine health information system currently is in the digital format. The data science approach also incorporates accountability and transparency which are now being realized as key issues when it comes to the use of health information. Reproducible algorithms can be used for revealing patterns of disease and transform health-related data for public health decision-making.

National Vector Borne Disease Control Programme (NVBDCP) is the nodal program for the prevention and control of vector-borne diseases in India (NVBDCP, 2021). Routine surveillance of vector-borne diseases is being carried out and data is generated from multiple health facilities. Dengue is a notifiable disease in the state of Punjab, India, and line listing of lab-confirmed cases are prepared for detailed epidemiological investigations, administrative requirements, and as decision support for the institution of preventive and control measures. Use of these routine health surveillance datasets to understand spatiotemporal patterns of dengue and linking with routine data from non-health sectors to understand determinants (climatic, environmental, socio-demography, health systems, etc.) will enable an in-depth understanding of dengue situation and development of disease forecasting models. This will strengthen existing surveillance mechanisms, and thus improve the health of the populations. However, for cross-linking of datasets, and the conduct of data analytics for knowledge generation, it is essential to clean the datasets in a manner that analysis-ready datasets are prepared from raw data without losing information. Thus, the present study was conducted to develop a rule-based reproducible and scalable logic model for cleaning routine health surveillance data in India using NVBDCP, Punjab program data as an illustrative example.

## 2. MATERIAL AND METHODS

2.1. **Data source.** Routine health care surveillance data provided by National Vector Borne Disease Control Programme, Directorate of Health Services, Government of Punjab, India. The datasets are composed of line listing data of lab-confirmed Dengue cases in the state from 01 January 2015 to 31 December 2019.

2.2. **Study variables.** The variables extracted from the routine data line list included information on the age of the patient, gender, place of occurrence, type of test performed, testing facility, and dates of testing, reporting, outpatient consultation, admission, and discharge.

2.3. **Framework.** The present study was conducted using the framework provided by Broeck et al. for data cleaning as a process (Van den Broeck et al., 2005). According to the framework, a data cleaning process is integral to all the components of a study process viz. study designing, data collection, data transformation, data extraction, data transfers, data exploration, and data analysis. The data cleaning process is a logical sequence of screening, diagnosing, and editing. The datasets are screened for anomalies and diagnosed to determine whether the anomaly is a true normal value, true extreme, an error, or undiagnosed with available data. This is followed by editing the data values by correction, deletion, or leaving as unchanged.

2.4. **Study design.** The data science approach was used for screening, diagnosis, and editing of raw datasets within the broad framework stated above. A reproducible algorithm was prepared which included a query code for screening, check code for diagnosis, and correction code for editing of data. Apriori definitions for valid data values expected range, data type, outliers, and data entry anomalies were created. The raw datasets were then systematically screened using the prepared algorithm for validity, presence of additional information, inconsistencies, strange patterns, and misplaced data in the line list. Once identified, the observed value, expected value, and neighborhood values were compared to confirm the presence of data issues that can be cleaned using the algorithm. All the data anomalies which were diagnosed to be due to apriori definitions were cleaned and data was extracted using an automated algorithm. The data anomalies wherein strange patterns were identified but had implicit valid values, a manual correction was carried out. In case of failure to obtain any valid value, the respective data cell was considered to be missing.

2.5. **Logic model**. A schematic representation of the logic model used for the data cleaning process is represented in Figure 1. The model imports the dataset and applies a rule-based, interactive algorithm to develop tidy data. The algorithm developed enquires about multiple possible anomalies in the dataset which can be cleaned in a semi-automated manner to extract information, and thus avoid loss of information for the respective variables. Each inquiry was based on the detection of string patterns in each reported case followed by an automated correction code on confirmation. The algorithm had subsets of inquiries for each variable and was run on all the cells in a phased manner for the data cleaning process. Data values were standardized by importing as text variables and creating a tidy string variable as the first step of the data cleaning process.

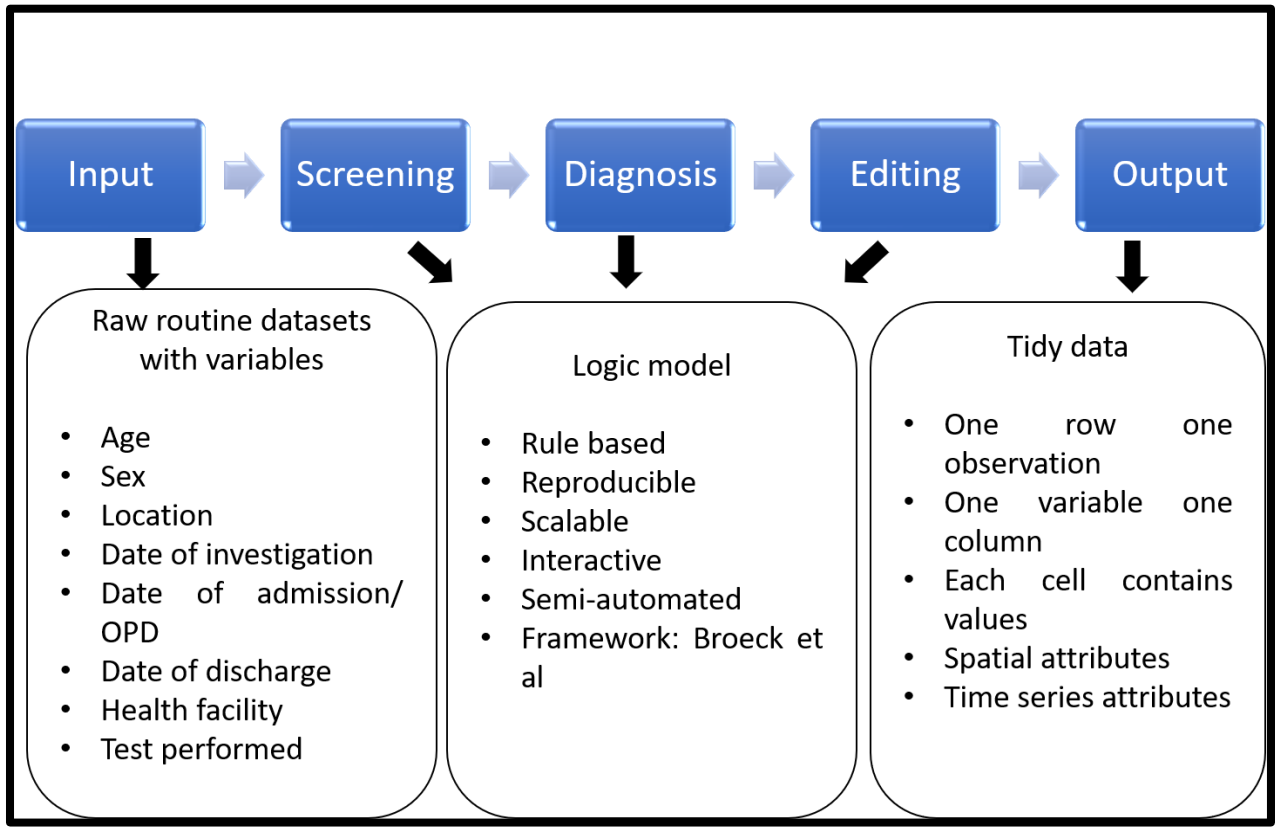

*Figure 1 Schematic representation of the data cleaning process.*

2.6. **Operational definitions and computational workflows.**

2.6.1. **Data cleaning process for date variables.** Date information can be analyzed when the date variable is present in a standard date format (e.g. ISO format). The date values were categorized into two types viz "excel-numeric" and "as-typed". Excel-numeric dates are values were defined as values with five characters, all digits, and no separator between digits. As-typed format values varied from a minimum length of 4 (e.g.: "2918" can represent for 02 September 2018) and had variations resulting from the field data entry personnel preferences (e.g.: "04 Jan 2020", "04-01-2020", "04/1/20", etc.). All the date values were read as string/text/character variables and standardized by removing all punctuations, separators, and whitespace. Each data value was then screened for the format of the date variable. In the case of the "excel-numeric" format, the date value was extracted by calculating the number of days since 01 January 1990. In the case of the "as typed" format, the value was screened for anomalies and data editing was carried out using the algorithm. All date values were transformed to the "ddmmyy" format for data extraction. Further, for missing data, data imputation by addition of mean days to testing from date of admission/OPD and subtracting mean days from discharge was carried out to estimate the date of testing.

2.6.2. **Data cleaning process to extract age-related information.** Age in analyzable format was defined as a numeric variable between 0 to 120 years. The expected column containing age details was imported as a text/ string/ character variable. All cells were screened for the presence of digit and character values. The values with the presence of non-digit characters were screened for the presence of valid digit values and data editing was carried out based on the findings. The alternate columns (e.g. sex details) which are likely to contain misplaced values for missing data cells in the age column were screened in a phased manner for enhancing the data extraction process.

2.6.3. **Data cleaning process to extract sex details**. The sex variable was defined as a factor variable with three levels viz. Male, Female, and Transgenders. All the cells were screened for the presence of non-case-sensitive keywords including "Male", "Female", "M" without "F", "F", "Child", "Transgender", and "TG". Data values containing digit characters were cleaned by deletion of digits and standardizing character values to lower case, removal of punctuations, and whitespaces. To obtain data on missing values, other columns were screened in a phased manner using keywords.

2.6.4. **Data cleaning process to extract location details**. The address variable was defined as a string/ text variable containing information related to the district, sub-district/ block, city, village, and town details. To extract location details, the addresses were standardized before bulk geocoding. Bulk geocoding was done using google API client services. The location details were extracted from the raw address using a reference list of blocks, cities, towns, and villages adapted from Census 2011 datasets. Information related to the patient/ caretaker was found to be present in raw address datasets, especially for the children during exploratory data analysis. The same was removed from the raw addresses through regular expressions-based text mining approaches.

2.6.5. **Data anonymization**. All the identifiers such as name and contact details were removed from the dataset. R package. *Epitrix* was used to generate anonymized data using the "scrypt" algorithm.

2.7. **Software.** All the data cleaning algorithms were prepared and executed in R software (R Core Team, 2020) using *tidyverse, stringr, lubridate, ggmaps, and epitrix* packages.

2.8. **Ethics statement**. The present study is part of a larger research project culminating in the Ph.D. program of the first author. Institutional Ethics Committee (IEC/IEC-1653; IEC Reg. No. ECR/189/Inst/KL/2013/RR-16) clearance obtained vide letter SCT/IEC/IEC-1653/DECEMBER-2020 dated 19/12/2020. Permission for use of program data has been obtained from the Directorate of Health Services, Government of Punjab, India.

3. **RESULTS**

The algorithm was executed for line listing data of 64,688 lab-confirmed dengue cases reported from the state during the study period. Logic algorithm for date extraction with screening results, automated cleaning process using the algorithm, and manual corrections are represented in Table 1. A total of 2,04,985 cells expected to contain date values were screened. The excel-numeric format was found in 42,902 cell values. Among 1,31,931 values identified by the screening algorithm, 1,31,569 (99.72%) were cleaned using the apriori cleaning codes (automated) after confirmation of screening results, and 362 (0.27%) values required manual correction.

| Logic algorithm | Rationale | Automated data cleaning process | Screening results | Automated | Manual correction |
|---|---|---|---|---|---|
| Total values screened | 204985 | | | | |
| Date | To covert string variable to date, same pattern should be present across the cells for correct parsing. In this pipeline, all the different date formats are converted to ddmmyyyy format for standardization of pattern and thus correct parsing of dates. | | | | |
| Presence of non-digit characters | Identify date specification using month values such as Jan, Feb, etc- | Remove all other non-digit characters. | 41747 | 41747 | 0 |
| Five-digit character values not starting with excel numeric date digit format and last two digits in yy format | All five-digit dmmyy/ddmyy format values will end in yy format. | Replace last two digits with yyyy year format. | 5627 | 5627 | 0 |
| Five-digit character values numerically outside the range of excel format dates and last two digits in yy format | Remaining five digit character values in as-typed format will be outside the range of number of days since 1899-12-30 to start and end days of the specified year. | Replace the last two digits with yyyy year format. | 202 | 202 | 0 |
| Five-digit character values numerically outside the range of excel format dates and not ending in yy format | Erroneous data values | Deletion. | 257 | 228 | 29 |
| Five-digit character values numerically within the range of excel format dates | Five-digit character values in excel format will be within the range of number of days since 1899-12-30 for the specified year | Excel format date extraction for five-digit values. | 42902 | 42902 | 0 |
| Any value with length more than eight characters. | Maximum length of dates in ddmmyyyy format is eight | Deletion. | 565 | 512 | 53 |
| **Eight-digit character values** | | | | | |
| Values not ending with yyyy format of the specified year | All eight-digit character dates should end with yyyy year format | Replace the last four digits with yyyy year format. | 168 | 119 | 49 |
| Month location holding a value greater than 12. | All eight-digit character dates should be ddmmyyyy format for parsing dates | Convert to ddmmyyyy format from mmddyyyy format. | 81 | 74 | 7 |
| **Seven-digit character values** | | | | | |
| Values ending digit as 1. | Mention of NS-1 positive along with dates in the raw data introduces error | Remove 1 from the last position. | 443 | 438 | 5 |
| Values not ending in yyyy format. | All seven-digit dates ending in yyyy format | Replace last four digits by 2019. | 157 | 72 | 85 |
| Values starting with 311, 211, or 111, and ending in the yyyy year format | Dates in seven-digit character have similar pattern on occurrence in month of January and November | If the date is in December or November, no changes are required. Else, replace the value with ddmmyyyy format. | 931 | 931 | 0 |
| Values with first two digits equal to | 1st, 2nd, and 3rd of every month till September and 10th, 20th and 30th have | If the dates are 1st, 2nd or 3rd, (dmmyyyy) no changes required. Else | 1181 | 1160 | 21 |
| 10 and ending in yyyy format. | same pattern in seven digit character format. | (ddmyyyy) insert a zero in the third location to create ddmmyyyy format. | | | |
| Values starting with zero. | Any seven-digit characters value starting cannot start with zero for parsing dates. | If the value is in ddmyyyy format, insert a zero at the third place. | 38 | 36 | 2 |
| Values ending in yyyy year format and numerically second and third location value is less than or equal to 12. | In seven-digit character values, from January to September, till 9th of every month (9122019), dates are written in dmmyyyy format. Then, from 11th to 31st of every month, dates are written in ddmyyyy format. | Insert a zero at first position to convert the value into ddmmyyyy format. | 5447 | 5447 | 0 |
| Remaining seven-digit character values | - | For values in ddmyyyy format, insert a zero at third location to convert it into ddmmyyyy format. | 6900 | 6887 | 13 |
| **Six-digit character values** | | | | | |
| Values starting with dmyyyy format. | Six-digit dates can be parsed and converted to date format when it is in ddmmyy format. In case of dmyyyy format or yyyydm format there will be error to parse. | Convert yyyydm into ddmmyyyy format. | 395 | 389 | 6 |
| Values which do not have last two digits as yy format. | All six-digit character values for a specified year should end in yy format | Replace last two digits with yy format. | 617 | 529 | 88 |
| Values which are ending in yy format, month location value is less than or equal to 12, and the date location value is less than or equal to 31 | Six-digit character values in ddmmyy format should be converted into ddmmyyyy format for similar pattern across dates for easy parsing at a later stage | Replace last two digits with yyyy format. | 23313 | 23310 | 3 |
| **Four-digit character values** | | | | | |
| Values with yyyy format | Four-digit character values should be in dmyy format | Convert to ddmmyyyy format or manual correction | 0 | 0 | 0 |
| Values not ending with yy format | Values with ddmm format | Convert to ddmmyyyy format or manual correction | 28 | 27 | 1 |

| Remaining four-digit character values | | All values in dmyy format to be converted to ddmmyyyy format by inserting additional zeros for day and month, and replacing yy with yyyy format, or manual correction | 307 | 307 | 0 |
|---|---|---|---|---|---|
| Values with three and less digit characters | | Deletion/ Manual correction | 625 | 625 | 0 |
| **Total** | | | 131931 | 131569 (99.72%) | 362 (0.27%) |

*Table 1 Logic model characteristics and performance for date extraction.*

Data extraction details for date, age, and sex variables from the dataset using the logic model are represented in **Figure 2**. The algorithm was able to clean and compute the estimated date of testing for 96.1% and 98.9% of observations for the year 2015 and 2016 respectively, and for all cases in the year 2017, 2018, and 2019. Age details were extracted maximum for the year 2017 and 2019 (99.4%) and minimum for the year 2015 (98.4%). Information on the sex of the patient was available for more than 99 percent across the study period, and location details were available for all the reported cases during the study period.

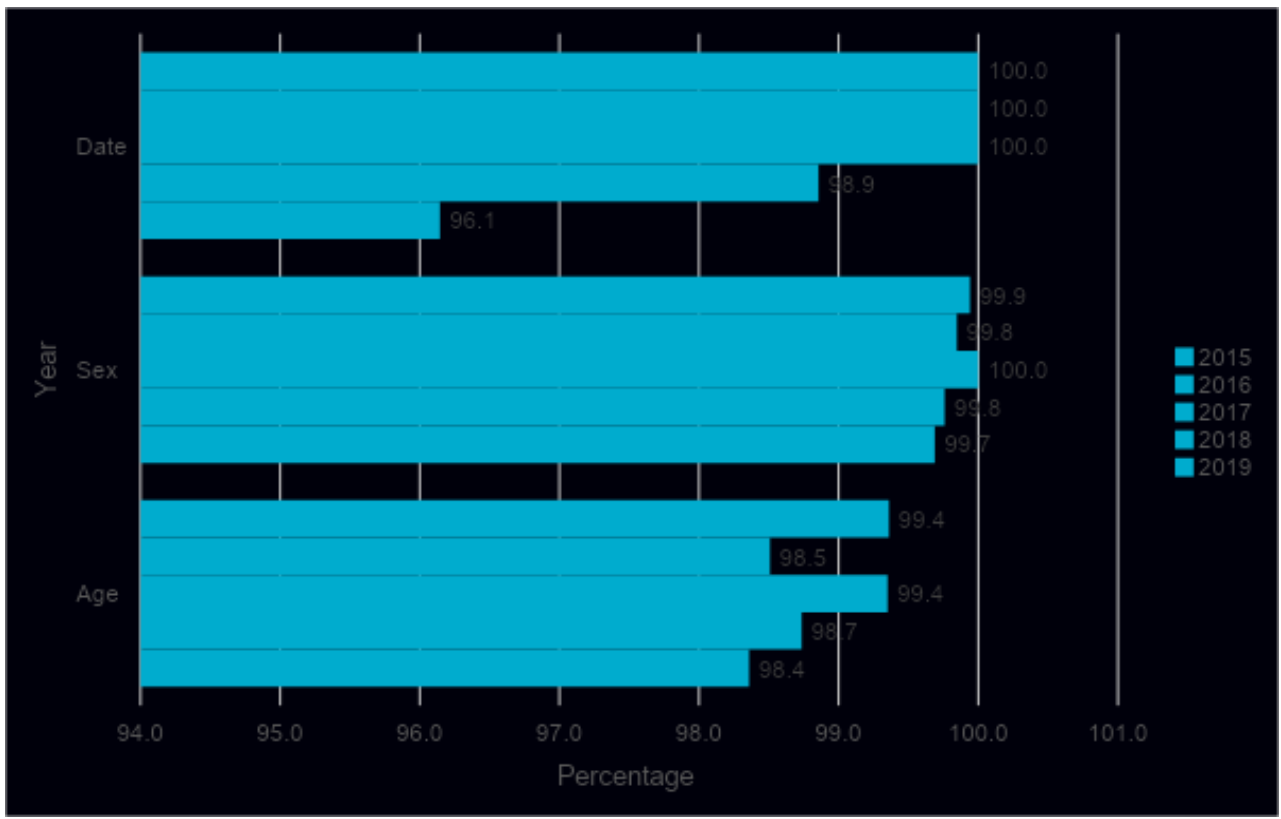

*Figure 2 Data extraction summary for age, gender, and estimated date of testing.*

## 4. DISCUSSION

The present study documents a systematic and reproducible logic model for data cleaning of routine health surveillance datasets in India. Though the systematic approach for data cleaning has been documented earlier on routine health information datasets (Gesicho et al., 2020; Maina et al., 2017; Phan et al., 2020), this study is novel in its application and illustration on routine program level dataset in India. Also, the present study was based on a data science approach that is increasingly being used for data analysis in epidemiology, but its utility in the development of reproducible and scalable data cleaning models has limited documentation. A recently published systematic review looking at the strategies applied in research articles to counter the issues of RHIS data quality across low- and middle-income countries suggest that majority of the studies that used RHIS data neither described the extent of the quality issues nor the steps they took to overcome them (Hung et al., 2020). The logic model developed in this study is expected to provide a practical strategy to clean routine health information program datasets in India resulting in strengthening of data quality for information and knowledge generation in the decision-making process as well as for research purposes.

The algorithm developed screened the data for date variables in a logical systematic approach. Among multiple variables present in the datasets, the timeline of disease occurrence is of utmost importance when analysis for disease patterns and model development is considered. Time series analysis is the most common analytical method followed by geostatistical analysis in routine data analytic studies (Hung et al., 2020). However, the dates are entered in varied formats in routine health information systems. This may be attributed to the use of basic data entry platforms such as Microsoft excel in the system. Though it suffices the "intended use" as defined for good quality data for day-to-day performance within the existing system, digital transformation of health care surveillance can be achieved by incorporating advancements in data handling and management technologies. Engagement of both data producers and users, identification of information needs, capacity building for data use at multiple levels, strengthening of data use and demand infrastructure are recommended measures for enhancement of data use context in health care systems (Nutley & Reynolds, 2013).

The present study used a rule-based semi-automated logic algorithm for data cleaning. Data cleaning approaches commonly used are broadly classified as logic-based and quantitative approaches. The use of Machine Learning and Artificial Intelligence based automated data cleaning workflows are largely based on the metadata of datasets. The semi-automated approach was chosen in the present context as it allows the user to understand the data along with the cleaning process in an iterative manner. The routine data currently in the country can be considered as digitalized as compared to the process of digital transformation wherein open data standards and metadata are inherent in database management systems. Initiatives for such database systems are required to enable the adoption of automated data cleaning workflows.

The presence of missing data values for the selected variables was found to be lower in the present study. This is in contrast to reported data missingness percentages in previous studies using routine datasets. This may be attributed to the type of variables selected and their perceived importance in the primary data use process. The line listing datasets prepared in the NVBDCP program include details on a limited number of variables that are considered essential for decision-making in the program. Further, the data values for a specified variable which seemed missing were found to be more commonly misplaced in the dataset. As a result, if alternate columns which are likely to hold information are not processed, the rate of missing data will be higher.

The research dissemination and uptake in health services require a collaborative approach between decision-makers and researchers to optimally utilize the advancements in information and digital technologies in health care. Data availability of program data for research purposes is required. Studies have proved that with increasing use of routine health data in decision making as well as for research purposes creates a self-perpetuating milieu in the data environment leading to improved data quality and strengthening of health systems.

Study limitations. Understanding the reasons behind the data anomalies present in the routine datasets is a critical factor to guide interventions to improve data quality. However, its understanding was beyond the scope of the present study. The algorithm developed in the present study was based on a single disease dataset from the national vector-borne disease control program. Its application in other diseases and program datasets may require additional screening mechanisms on one hand and may not require some screening steps on the other. Future studies on the application of the algorithm for external generalizability will establish the robustness of the algorithm for larger use. Similarly, limited variables required for the present project were explored in the present study, however, being scalable, algorithms for additional variables as required by health program managers and for research purposes can be incorporated into the model.

The strength of the present study includes the use of a reproducible and scalable logic algorithm for data preprocessing of routine health surveillance data. This will enable the researchers to start looking at available routine datasets. The resulting dataset can be used to understand the Spatio-temporal epidemiology of diseases. Good quality data can be used to develop forecasting models which can complement existing surveillance mechanisms and reduce disease-related burden in the populations. The scalability of algorithms prepared in open-source software provides enormous potential for application to routine datasets for other diseases and for geographical regions with similar challenges globally. However, the conceptualization of development in ICT4D involves careful understanding of the research context within the broader goals for sustainable development. The institutionalization mechanisms for outcomes of ICT4D research will require ingrained perspectives related to the dimensions and theories of change for development (Zheng et al., 2018).

## 5. CONCLUSION

Data quality of routine health information systems can be strengthened using systematic, reproducible algorithms for data cleaning in open-source software. The algorithm in the present study was semi-automated and based on routine health surveillance data in India. It resulted in the development of a research-level dataset that can be analyzed and interlinked with data from non-health sectors, thus illuminating one of the key contributions of data science to public health systems. Being scalable, the implication of this information and digital technology in health systems and digital epidemiology is enormous. The logic model can be expanded for additional variables according to the health system and research needs in the future.

## REFERENCES AND CITATIONS